\documentclass[aps,nofootinbib,preprint,superscriptaddress]{revtex4}%
\usepackage{hyperref}
\usepackage{amsmath}
\usepackage{amsfonts}
\usepackage{amssymb}
\usepackage{graphicx}
\usepackage{color}%
\providecommand{\U}[1]{\protect\rule{.1in}{.1in}}

\begin{document}
\title{The SL(K+3,C) Symmetry of the Bosonic String Scattering Amplitudes}
\author{Sheng-Hong Lai}
\email{xgcj944137@gmail.com}
\affiliation{Department of Electrophysics, National Chiao-Tung University, Hsinchu, Taiwan, R.O.C.}
\author{Jen-Chi Lee}
\email{jcclee@cc.nctu.edu.tw}
\affiliation{Department of Electrophysics, National Chiao-Tung University, Hsinchu, Taiwan, R.O.C.}
\author{Yi Yang}
\email{yiyang@mail.nctu.edu.tw}
\affiliation{Department of Electrophysics, National Chiao-Tung University, Hsinchu, Taiwan, R.O.C.}
\author{}
\date{\today }

\begin{abstract}
We discover that the exact string scattering amplitudes (SSA) of three
tachyons and one arbitrary string state, or the Lauricella SSA (LSSA), in the
$26D$ open bosonic string theory can be expressed in terms of the basis
functions in the infinite dimensional representation space of the $SL(K+3,%
\mathbb{C}
)$ group. In addition, we find that the $K+2$ recurrence relations among the
LSSA discovered by the present authors previously can be used to reproduce the
Cartan subalgebra and simple root system of the $SL(K+3,%
\mathbb{C}
)$ group with rank $K+2$. As a result, the $SL(K+3,%
\mathbb{C}
)$ group can be used to solve all the LSSA and express them in terms of one
amplitude. As an application in the hard scattering limit, the $SL(K+3,%
\mathbb{C}
)$ group can be used to directly prove Gross conjecture
\cite{GM,Gross,GrossManes}, which was previously corrected and proved by the
method of decoupling of zero norm states \cite{ChanLee1,ChanLee2,CHL,PRL,
CHLTY,susy,ZNS1}.

\end{abstract}
\maketitle
\tableofcontents

%

\setcounter{equation}{0}
\renewcommand{\theequation}{\arabic{section}.\arabic{equation}}%

\section{Introduction}

One of the most important issue of string theory is its spacetime symmetry
structure. It has been widely believed that there exist huge spacetime
symmetries of string theory. One way to study string symmetry is to calculate
string scattering amplitudes (SSA). Indeed, it was conjectured by Gross
\cite{GM,Gross,GrossManes} that there exist infinite number of linear
relations among high energy, fixed angle or hard SSA of different string
states. This conjecture was later corrected and explicitly proved in
\cite{ChanLee1,ChanLee2,CHL,PRL, CHLTY,susy} by using the method of decoupling
of zero-norm states \cite{ZNS1}. Moreover, these infinite linear relations are
so powerful that they can be used to reduce the number of independent hard SSA
from $\infty$ down to $1$. Other approaches of stringy symmetries can be found
at \cite{Moore,Moore1,CKT,Sag}. For more details, see \cite{review} for a
recent review.

On the other hand, it was found that the high energy, fixed momentum transfer
or Regge SSA of three tachyons and one arbitrary string states can be
expressed in terms of a sum of Kummer functions $U$ \cite{KLY,LY,LY2014},
which were then shown to be the first Appell function $F_{1}$ \cite{LY2014}.
Regge stringy recurrence relations \cite{LY,LY2014} can then be constructed
and used to reduce the number of independent\ Regge SSA from $\infty$ down to
$1$. Moreover, an interesting link between Regge SSA and hard SSA was pointed
out in \cite{KLY,LYY}, and for each mass level the ratios among hard SSA can
be extracted from Regge SSA. It was then conjectured that the $SL(5;C)$
dynamical symmetry of the Appell function $F_{1}$ \cite{sl5c} is crucial to
probe high energy spacetime symmetry of string theory.

More recently, the Lauricella string scattering amplitudes (LSSA) \cite{LLY2}
of three tachyons and one arbitrary string state in the $26D$ open bosonic
string theory valid for \textit{arbitrary energies} were calculated and
expressed in terms of the D-type Lauricella functions $F_{D}^{(K)}.$ Moreover,
it was shown that \cite{LLLY} there exist $K+2$ recurrence relations among
$F_{D}^{(K)}$ which (together with a multiplication theorem of $F_{D}^{(1)}$)
can be used to derive recurrence relations among LSSA and reduce the number of
independent\ LSSA from $\infty$ down to $1$.

In this paper, we will show the existence of the spacetime symmetry group
structure of the LSSA. To be more specific, we will demonstrate that the LSSA
can be expressed in terms of the basis functions in the infinite dimensional
representation space of the $SL(K+3,%
\mathbb{C}
)$ group \cite{sl4c,slkc} which contains the $SO(2,1)$ spacetime Lorentz
group. In addition, we find that the $K+2$ recurrence relations among the LSSA
discovered by the present authors \cite{LLLY} previously can be used to
reproduce the Cartan subalgebra and simple root system of the $SL(K+3,%
\mathbb{C}
)$ group with rank $K+2$.

We thus have demonstrated the existence of a spacetime symmetry group of the
$26D$ open bosonic string scattering amplitudes. As a result, the $SL(K+3,%
\mathbb{C}
)$ group can be used to solve all the LSSA and express them in terms of one
amplitude. As an application in the hard scattering limit, the $SL(K+3,%
\mathbb{C}
)$ group can be used to directly prove Gross conjecture
\cite{GM,Gross,GrossManes}, which was previously corrected and proved by the
method of decoupling of zero norm states \cite{ChanLee1,ChanLee2,CHL,PRL,
CHLTY,susy,ZNS1}.%

\setcounter{equation}{0}
\renewcommand{\theequation}{\arabic{section}.\arabic{equation}}%

\section{Review of the LSSA}

In this section, we first review the LSSA of three tachyons and one arbitrary
string states of the 26D open bosonic string. The general states at mass level
$M_{2}^{2}=2(N-1)$, $N=\sum_{n,m,l>0}\left(  nr_{n}^{T}+mr_{m}^{P}+lr_{l}%
^{L}\right)  $ with polarizations on the scattering plane are of the form (see
below for the definition of polarizations of $T$, $L$ and $P$)%
\begin{equation}
\left\vert r_{n}^{T},r_{m}^{P},r_{l}^{L}\right\rangle =\prod_{n>0}\left(
\alpha_{-n}^{T}\right)  ^{r_{n}^{T}}\prod_{m>0}\left(  \alpha_{-m}^{P}\right)
^{r_{m}^{P}}\prod_{l>0}\left(  \alpha_{-l}^{L}\right)  ^{r_{l}^{L}}%
|0,k\rangle. \label{state}%
\end{equation}
In the CM frame, the kinematics are defined as%
\begin{align}
k_{1}  &  =\left(  \sqrt{M_{1}^{2}+|\vec{k_{1}}|^{2}},-|\vec{k_{1}}|,0\right)
,\\
k_{2}  &  =\left(  \sqrt{M_{2}+|\vec{k_{1}}|^{2}},+|\vec{k_{1}}|,0\right)  ,\\
k_{3}  &  =\left(  -\sqrt{M_{3}^{2}+|\vec{k_{3}}|^{2}},-|\vec{k_{3}}|\cos
\phi,-|\vec{k_{3}}|\sin\phi\right)  ,\\
k_{4}  &  =\left(  -\sqrt{M_{4}^{2}+|\vec{k_{3}}|^{2}},+|\vec{k_{3}}|\cos
\phi,+|\vec{k_{3}}|\sin\phi\right)
\end{align}
with $M_{1}^{2}=M_{3}^{2}=M_{4}^{2}=-2$ and $\phi$ is the scattering angle.
The Mandelstam variables are $s=-\left(  k_{1}+k_{2}\right)  ^{2}$,
$t=-\left(  k_{2}+k_{3}\right)  ^{2}$ and $u=-\left(  k_{1}+k_{3}\right)
^{2}$. There are three polarizations on the scattering plane
\cite{ChanLee1,ChanLee2}%
\begin{align}
e^{T}  &  =(0,0,1),\\
e^{L}  &  =\frac{1}{M_{2}}\left(  |\vec{k_{1}}|,\sqrt{M_{2}+|\vec{k_{1}}|^{2}%
},0\right)  ,\\
e^{P}  &  =\frac{1}{M_{2}}\left(  \sqrt{M_{2}+|\vec{k_{1}}|^{2}},|\vec{k_{1}%
}|,0\right)
\end{align}
where $e^{P}=\frac{1}{M_{2}}(E_{2},\mathrm{k}_{2},0)=\frac{k_{2}}{M_{2}}$ the
momentum polarization, $e^{L}=\frac{1}{M_{2}}(\mathrm{k}_{2},E_{2},0)$ the
longitudinal polarization and $e^{T}=(0,0,1)$ the transverse polarization. For
later use, we define%
\begin{equation}
k_{i}^{X}\equiv e^{X}\cdot k_{i}\text{ \ for \ }X=\left(  T,P,L\right)  .
\end{equation}
It is important to note that SSA of three tachyons and one arbitrary string
state with polarizations orthogonal to the scattering plane vanish. Thus the
Lorentz spacetime symmetry group is $SO(2,1)$. The $\left(  s,t\right)  $
channel of the LSSA can be calculated to be \cite{LLY2}%
\begin{align}
A_{st}^{(r_{n}^{T},r_{m}^{P},r_{l}^{L})}  &  =\prod_{n=1}\left[
-(n-1)!k_{3}^{T}\right]  ^{r_{n}^{T}}\cdot\prod_{m=1}\left[  -(m-1)!k_{3}%
^{P}\right]  ^{r_{m}^{P}}\prod_{l=1}\left[  -(l-1)!k_{3}^{L}\right]
^{r_{l}^{L}}\nonumber\\
&  \cdot B\left(  -\frac{t}{2}-1,-\frac{s}{2}-1\right)  F_{D}^{(K)}\left(
-\frac{t}{2}-1;R_{n}^{T},R_{m}^{P},R_{l}^{L};\frac{u}{2}+2-N;\tilde{Z}_{n}%
^{T},\tilde{Z}_{m}^{P},\tilde{Z}_{l}^{L}\right)  \label{st}%
\end{align}
where we have defined $R_{k}^{X}\equiv\left\{  -r_{1}^{X}\right\}  ^{1}%
,\cdots,\left\{  -r_{k}^{X}\right\}  ^{k}$ with $\left\{  a\right\}
^{n}=\underset{n}{\underbrace{a,a,\cdots,a}}$, $Z_{k}^{X}\equiv\left[
z_{1}^{X}\right]  ,\cdots,\left[  z_{k}^{X}\right]  $ with $\left[  z_{k}%
^{X}\right]  =z_{k0}^{X},\cdots,z_{k\left(  k-1\right)  }^{X}$ and $z_{k}%
^{X}=\left\vert \left(  -\frac{k_{1}^{X}}{k_{3}^{X}}\right)  ^{\frac{1}{k}%
}\right\vert $,\ $z_{kk^{\prime}}^{X}=z_{k}^{X}e^{\frac{2\pi ik^{\prime}}{k}}%
$,\ $\tilde{z}_{kk^{\prime}}^{X}\equiv1-z_{kk^{\prime}}^{X}$ for $k^{\prime
}=0,\cdots,k-1$. The integer $K$ in Eq.(\ref{st}) is defined to be%
\begin{equation}
\text{ }K=\underset{\{\text{for all }r_{j}^{T}\neq0\}}{\sum j}%
+\underset{\{\text{for all }r_{j}^{P}\neq0\}}{\sum j}+\underset{\{\text{for
all }r_{j}^{L}\neq0\}}{\sum j}.
\end{equation}
The D-type Lauricella function $F_{D}^{(K)}$ is one of the four extensions of
the Gauss hypergeometric function to $K$ variables and is defined as%
\begin{align}
&  F_{D}^{(K)}\left(  \alpha;\beta_{1},...,\beta_{K};\gamma;x_{1}%
,...,x_{K}\right) \nonumber\\
&  =\sum_{n_{1},\cdots,n_{K}=0}^{\infty}\frac{\left(  \alpha\right)
_{n_{1}+\cdots+n_{K}}}{\left(  \gamma\right)  _{n_{1}+\cdots+n_{K}}}%
\frac{\left(  \beta_{1}\right)  _{n_{1}}\cdots\left(  \beta_{K}\right)
_{n_{K}}}{n_{1}!\cdots n_{K}!}x_{1}^{n_{1}}\cdots x_{K}^{n_{K}}%
\end{align}
where $(\alpha)_{n}=\alpha\cdot\left(  \alpha+1\right)  \cdots\left(
\alpha+n-1\right)  $ is the Pochhammer symbol. There was a integral
representation of the Lauricella function $F_{D}^{(K)}$ discovered by Appell
and Kampe de Feriet (1926) \cite{Appell}%
\begin{align}
&  F_{D}^{(K)}\left(  \alpha;\beta_{1},...,\beta_{K};\gamma;x_{1}%
,...,x_{K}\right) \nonumber\\
&  =\frac{\Gamma(\gamma)}{\Gamma(\alpha)\Gamma(\gamma-\alpha)}\int_{0}%
^{1}dt\,t^{\alpha-1}(1-t)^{\gamma-\alpha-1}\cdot(1-x_{1}t)^{-\beta_{1}%
}(1-x_{2}t)^{-\beta_{2}}...(1-x_{K}t)^{-\beta_{K}}, \label{Kam}%
\end{align}
which was used to calculate Eq.(\ref{st}).

\bigskip To illustrate the complicated notations used in Eq.(\ref{st}), we
give three explicit examples of the LSSA in the following.

\subsection{Example one}

We take the tensor state to be%
\begin{equation}
\left\vert \text{state}\right\rangle =\left(  \alpha_{-1}^{T}\right)
^{r_{1}^{T}}\left(  \alpha_{-1}^{P}\right)  ^{r_{1}^{P}}\left(  \alpha
_{-1}^{L}\right)  ^{r_{1}^{L}}|0,k\rangle.
\end{equation}
The LSSA in Eq.(\ref{st}) can be calculated to be%
\begin{align}
A_{st}^{(r_{1}^{T},r_{1}^{P},r_{l}^{L})}  &  =\left(  -k_{3}^{T}\right)
^{r_{1}^{T}}\left(  -k_{3}^{P}\right)  ^{r_{1}^{P}}\left(  -k_{3}^{L}\right)
^{r_{1}^{L}}B\left(  -\frac{t}{2}-1,-\frac{s}{2}-1\right) \nonumber\\
&  \cdot F_{D}^{(3)}\left(  -\frac{t}{2}-1;-r_{1}^{T},-r_{1}^{P},-r_{1}%
^{L};\frac{u}{2}+2-N;\tilde{z}_{10}^{T},\tilde{z}_{10}^{P},\tilde{z}_{10}%
^{L}\right)
\end{align}
where the arguments in $F_{D}^{(3)}$ are calculated to be%
\begin{align}
R_{n}^{T}  &  =\left\{  -r_{1}^{T}\right\}  ^{1},\cdots,\left\{  -r_{n}%
^{T}\right\}  ^{k}=\left\{  -r_{1}^{T}\right\}  ^{1}=-r_{1}^{T},\nonumber\\
R_{m}^{P}  &  =\left\{  -r_{1}^{P}\right\}  ^{1},\cdots,\left\{  -r_{m}%
^{P}\right\}  ^{k}=\left\{  -r_{1}^{P}\right\}  ^{1}=-r_{1}^{P},\nonumber\\
R_{l}^{L}  &  =\left\{  -r_{1}^{L}\right\}  ^{1},\cdots,\left\{  -r_{l}%
^{L}\right\}  ^{k}=\left\{  -r_{1}^{L}\right\}  ^{1}=-r_{1}^{L},\\
\tilde{Z}_{n}^{T}  &  =\left[  \tilde{z}_{1}^{T}\right]  ,\cdots,\left[
\tilde{z}_{n}^{T}\right]  =\left[  \tilde{z}_{1}^{T}\right]  =\tilde{z}%
_{10}^{T}=1-z_{10}^{T}=1-z_{k}^{T}e^{\frac{2\pi i0}{1}}=1-\left\vert
-\frac{k_{1}^{T}}{k_{3}^{T}}\right\vert ,\nonumber\\
\tilde{Z}_{n}^{P}  &  =\left[  \tilde{z}_{1}^{P}\right]  ,\cdots,\left[
\tilde{z}_{n}^{P}\right]  =\left[  \tilde{z}_{1}^{P}\right]  =\tilde{z}%
_{10}^{P}=1-\left\vert -\frac{k_{1}^{P}}{k_{3}^{P}}\right\vert ,\nonumber\\
\tilde{Z}_{n}^{L}  &  =\left[  \tilde{z}_{1}^{L}\right]  ,\cdots,\left[
\tilde{z}_{n}^{L}\right]  =\left[  \tilde{z}_{1}^{L}\right]  =\tilde{z}%
_{10}^{L}=1-\left\vert -\frac{k_{1}^{L}}{k_{3}^{L}}\right\vert
\end{align}
and%
\begin{align}
\text{ }K  &  =\underset{\{\text{for all }r_{j}^{T}\neq0\}}{\sum
j}+\underset{\{\text{for all }r_{j}^{P}\neq0\}}{\sum j}+\underset{\{\text{for
all }r_{j}^{L}\neq0\}}{\sum j}\nonumber\\
&  =1+1+1=3.
\end{align}

\subsection{Example two}

We take the tensor state to be%
\begin{equation}
\left\vert \text{state}\right\rangle =\left(  \alpha_{-1}^{T}\right)
^{r_{1}^{T}}\left(  \alpha_{-2}^{T}\right)  ^{r_{2}^{T}}|0,k\rangle.
\end{equation}
The LSSA in Eq.(\ref{st}) can be calculated to be%
\begin{align}
A_{st}^{(r_{1}^{T},r_{1}^{P},r_{l}^{L})}  &  =\left(  -k_{3}^{T}\right)
^{r_{1}^{T}}\left(  -k_{3}^{T}\right)  ^{r_{2}^{T}}B\left(  -\frac{t}%
{2}-1,-\frac{s}{2}-1\right) \nonumber\\
&  \cdot F_{D}^{(3)}\left(  -\frac{t}{2}-1;-r_{1}^{T},-r_{2}^{T},-r_{2}%
^{T};\frac{u}{2}+2-N;\tilde{z}_{10}^{T},\tilde{z}_{20}^{T},\tilde{z}_{21}%
^{T}\right)
\end{align}
where the arguments in $F_{D}^{(3)}$ are calculated to be%

\begin{align}
R_{n}^{T}  &  =\left\{  -r_{1}^{T}\right\}  ^{1},\cdots,\left\{  -r_{n}%
^{T}\right\}  ^{k}=\left\{  -r_{1}^{T}\right\}  ^{1},\left\{  -r_{2}%
^{T}\right\}  ^{2}=-r_{1}^{T},\underset{2}{\underbrace{-r_{2}^{T},-r_{2}^{T}}%
}\\
\tilde{Z}_{n}^{T}  &  =\left[  \tilde{z}_{1}^{T}\right]  ,\cdots,\left[
\tilde{z}_{n}^{T}\right]  =\left[  \tilde{z}_{1}^{T}\right]  ,\left[
\tilde{z}_{2}^{T}\right]  =\tilde{z}_{10}^{T},\tilde{z}_{20}^{T},\tilde
{z}_{21}^{T}\nonumber\\
&  =1-z_{10}^{T},1-z_{20}^{T},1-z_{21}^{T}\nonumber\\
&  =1-z_{1}^{T}e^{\frac{2\pi i0}{1}},1-z_{2}^{T}e^{\frac{2\pi i0}{2}}%
,1-z_{2}^{T}e^{\frac{2\pi i1}{2}}\nonumber\\
&  =1-z_{1}^{T},1-z_{2}^{T},1+z_{2}^{T}%
\end{align}
and%
\begin{align}
\text{ }K  &  =\underset{\{\text{for all }r_{j}^{T}\neq0\}}{\sum
j}+\underset{\{\text{for all }r_{j}^{P}\neq0\}}{\sum j}+\underset{\{\text{for
all }r_{j}^{L}\neq0\}}{\sum j}\\
&  =(1+2)+0+0=3.
\end{align}

\subsection{Example three}

We take the tensor state to be%
\begin{equation}
\left\vert \text{state}\right\rangle =\left(  \alpha_{-1}^{T}\right)
^{r_{1}^{T}}\left(  \alpha_{-2}^{T}\right)  ^{r_{2}^{T}}\left(  \alpha
_{-5}^{T}\right)  ^{r_{5}^{T}}\left(  \alpha_{-6}^{T}\right)  ^{r_{6}^{T}%
}|0,k\rangle.
\end{equation}
The LSSA in Eq.(\ref{st}) can be calculated to be%
\begin{align}
A_{st}^{(r_{1}^{T},r_{1}^{P},r_{l}^{L})}  &  =\left(  -k_{3}^{T}\right)
^{r_{1}^{T}}\left(  -k_{3}^{T}\right)  ^{r_{2}^{T}}\left(  -4!k_{3}%
^{T}\right)  ^{r_{5}^{T}}\left(  -5!k_{3}^{T}\right)  ^{r_{6}^{T}}B\left(
-\frac{t}{2}-1,-\frac{s}{2}-1\right) \nonumber\\
&  \cdot F_{D}^{(14)}\left(
\begin{array}
[c]{c}%
-\frac{t}{2}-1;-r_{1}^{T},\underset{2}{\underbrace{-r_{2}^{T},-r_{2}^{T}}%
},\underset{5}{\underbrace{-r_{5}^{T},-r_{5}^{T},-r_{5}^{T},-r_{5}^{T}%
,-r_{5}^{T}}},\underset{6}{\underbrace{-r_{6}^{T},-r_{6}^{T},-r_{6}^{T}%
,-r_{6}^{T},-r_{6}^{T},-r_{6}^{T}}};\\
\frac{u}{2}+2-N;\tilde{z}_{10}^{T},\underset{2}{\underbrace{\tilde{z}_{20}%
^{T},\tilde{z}_{21}^{T}}},\underset{5}{\underbrace{\tilde{z}_{50}^{T}%
,\tilde{z}_{51}^{T},\tilde{z}_{52}^{T},\tilde{z}_{53}^{T},\tilde{z}_{54}^{T}}%
},\underset{6}{\underbrace{\tilde{z}_{60}^{T},\tilde{z}_{61}^{T},\tilde
{z}_{62}^{T},\tilde{z}_{63}^{T},\tilde{z}_{64}^{T},\tilde{z}_{65}^{T}}}%
\end{array}
\right)
\end{align}
\newline where the arguments in $F_{D}^{(14)}$ are calculated to be%

\begin{align}
R_{n}^{T}  &  =\left\{  -r_{1}^{T}\right\}  ^{1},\cdots,\left\{  -r_{n}%
^{T}\right\}  ^{k}=\left\{  -r_{1}^{T}\right\}  ^{1},\left\{  -r_{2}%
^{T}\right\}  ^{2},\left\{  -r_{5}^{T}\right\}  ^{5},\left\{  -r_{6}%
^{T}\right\}  ^{6}\nonumber\\
&  =-r_{1}^{T},\underset{2}{\underbrace{-r_{2}^{T},-r_{2}^{T}}}%
,\underset{5}{\underbrace{-r_{5}^{T},-r_{5}^{T},-r_{5}^{T},-r_{5}^{T}%
,-r_{5}^{T}}},\underset{6}{\underbrace{-r_{6}^{T},-r_{6}^{T},-r_{6}^{T}%
,-r_{6}^{T},-r_{6}^{T},-r_{6}^{T}}}\\
\tilde{Z}_{n}^{T}  &  =\left[  \tilde{z}_{1}^{T}\right]  ,\cdots,\left[
\tilde{z}_{n}^{T}\right]  =\left[  \tilde{z}_{1}^{T}\right]  ,\left[
\tilde{z}_{2}^{T}\right]  ,\left[  \tilde{z}_{5}^{T}\right]  ,\left[
\tilde{z}_{6}^{T}\right] \nonumber\\
&  =\tilde{z}_{10}^{T},\underset{2}{\underbrace{\tilde{z}_{20}^{T},\tilde
{z}_{21}^{T}}},\underset{5}{\underbrace{\tilde{z}_{50}^{T},\tilde{z}_{51}%
^{T},\tilde{z}_{52}^{T},\tilde{z}_{53}^{T},\tilde{z}_{54}^{T}}}%
,\underset{6}{\underbrace{\tilde{z}_{60}^{T},\tilde{z}_{61}^{T},\tilde{z}%
_{62}^{T},\tilde{z}_{63}^{T},\tilde{z}_{64}^{T},\tilde{z}_{65}^{T}}}%
\end{align}
and%
\begin{align}
\text{ }K  &  =\underset{\{\text{for all }r_{j}^{T}\neq0\}}{\sum
j}+\underset{\{\text{for all }r_{j}^{P}\neq0\}}{\sum j}+\underset{\{\text{for
all }r_{j}^{L}\neq0\}}{\sum j}\nonumber\\
&  =\left(  1+2+5+6\right)  +0+0=14.
\end{align}

As an application of Eq.(\ref{st}), it can be shown that in the hard
scattering limit $e^{P}=e^{L}$ \cite{ChanLee1,ChanLee2}, the leading order
LSSA corresponds to $r_{1}^{T}=N-2m-2q$, $r_{1}^{L}=2m$ and $r_{2}^{L}=q$, and
the LSSA in the hard scattering limit can be calculated to be \cite{LLY2}
\begin{align}
A_{st}^{(N-2m-2q,2m,q)}  &  =B\left(  -\frac{t}{2}-1,-\frac{s}{2}-1\right)
\left(  E\sin\phi\right)  ^{N}\frac{\left(  2m\right)  !}{m!}\left(  -\frac
{1}{2M_{2}}\right)  ^{2m+q}\nonumber\\
&  =(2m-1)!!\left(  -\frac{1}{M_{2}}\right)  ^{2m+q}\left(  \frac{1}%
{2}\right)  ^{m+q}A_{st}^{(N,0,0)},
\end{align}
which gives the ratios \cite{review}
\begin{equation}
\frac{A_{st}^{(N-2m-2q,2m,q)}}{A_{st}^{(N,0,0)}}=(2m-1)!!\left(  -\frac
{1}{M_{2}}\right)  ^{2m+q}\left(  \frac{1}{2}\right)  ^{m+q}, \label{ratio}%
\end{equation}
and is consistent with the previous result \cite{ChanLee1,ChanLee2,CHL,PRL,
CHLTY,susy}. The first example calculated was the ratios at mass level
$M^{2}=4$ \cite{ChanLee1,ChanLee2}
\begin{equation}
\mathcal{T}_{TTT}:\mathcal{T}_{LLT}:\mathcal{T}_{(LT)}:\mathcal{T}%
_{[LT]}=8:1:-1:-1. \label{03}%
\end{equation}
The ratios among SSA in Eq.(\ref{ratio}) and Eq.(\ref{03}) are generalization
of ratios among field theory scattering amplitudes. Let's consider a simple
analogy from particle physics. The ratios of the nucleon-nucleon scattering
processes%
\begin{align}
(a)\text{ \ }p+p  &  \rightarrow d+\pi^{+},\nonumber\\
(b)\text{ \ }p+n  &  \rightarrow d+\pi^{0},\nonumber\\
(c)\text{ \ }n+n  &  \rightarrow d+\pi^{-} \label{05}%
\end{align}
can be calculated to be (ignore the tiny mass difference between proton and
neutron)%
\begin{equation}
T_{a}:T_{b}:T_{c}=1:\frac{1}{\sqrt{2}}:1 \label{06}%
\end{equation}
from $SU(2)$ isospin symmetry. Is there any symmetry group structure which can
be used to calculate SSA ratios in Eq.(\ref{ratio}) and Eq.(\ref{03})? This is
the main issue we want to address in this paper and it turns out that the
relevant group is the noncompact $SL(K+3,C)$ group as we will discuss in the
rest of the paper. Since the spacetime symmetry group of the LSSA needs to
include the noncompact Lorentz group $SO(2,1)$, the noncompact $SL(K+3,C)$
group seems to be a reasonable one.%

\setcounter{equation}{0}
\renewcommand{\theequation}{\arabic{section}.\arabic{equation}}%

\section{The SL(4,C) Symmetry}

In this section, for illustration we first consider the simplest $K=1$ case
with $SL(4,C)$ symmetry. For a given $K$, there can be LSSA with different
mass level $N$. For illustration, for $K=1$ as an example there are three
types of LSSA%
\begin{align}
(\alpha_{-1}^{T})^{p_{1}}\text{ , }F_{D}^{(1)}\left(  -\frac{t}{2}%
-1,-p_{1},,\frac{u}{2}+2-p_{1},1\right)  \text{ , }N  &  =p_{1},\nonumber\\
(\alpha_{-1}^{P})^{q_{1}}\text{ , }F_{D}^{(1)}\left(  -\frac{t}{2}%
-1,-q_{1},\frac{u}{2}+2-q_{1},\left[  \tilde{z}_{1}^{P}\right]  \right)
\text{ , }N  &  =q_{1},\nonumber\\
(\alpha_{-1}^{L})^{r_{1}}\text{ , }F_{D}^{(1)}\left(  -\frac{t}{2}%
-1,-r_{1},\frac{u}{2}+2-r_{1},\left[  \tilde{z}_{1}^{L}\right]  \right)
\text{ , }N  &  =r_{1}. \label{rel}%
\end{align}
To calculate the group representation of the LSSA for $K=1$, we first define
\cite{slkc}%

\begin{equation}
f_{ac}^{b}\left(  \alpha;\beta;\gamma;x\right)  =B\left(  \gamma-\alpha
,\alpha\right)  F_{D}^{\left(  1\right)  }\left(  \alpha;\beta;\gamma
;x\right)  a^{\alpha}b^{\beta}c^{\gamma}. \label{id}%
\end{equation}
Note that the LSSA in Eq.(\ref{st}) for $K=1$ corresponds to the case $a=1=c$,
and can be written as%
\begin{equation}
A_{st}^{R^{X}}=f_{11}^{-k_{3}^{X}}\left(  -\frac{t}{2}-1;R^{X};\frac{u}%
{2}+2-N;\tilde{Z}^{X}\right)  .
\end{equation}
We are now ready to introduce the $15$ generators of $SL(4,C)$ group
\cite{sl4c,slkc}%
\begin{align}
E_{\alpha}  &  =a\left(  x\partial_{x}+a\partial_{a}\right)  ,\nonumber\\
E_{-\alpha}  &  =\frac{1}{a}\left[  x\left(  1-x\right)  \partial
_{x}+c\partial_{c}-a\partial_{a}-xb\partial_{b}\right]  ,\nonumber\\
E_{\beta}  &  =b\left(  x\partial_{x}+b\partial_{b}\right)  ,\nonumber\\
E_{-\beta}  &  =\frac{1}{b}\left[  x\left(  1-x\right)  \partial_{x}%
+c\partial_{c}-b\partial_{b}-xa\partial_{a}\right]  ,\nonumber\\
E_{\gamma}  &  =c\left[  \left(  1-x\right)  \partial_{x}+c\partial
_{c}-a\partial_{a}-b\partial_{b}\right]  ,\nonumber\\
E_{-\gamma}  &  =-\frac{1}{c}\left(  x\partial_{x}+c\partial_{c}-1\right)
,\nonumber\\
E_{\beta\gamma}  &  =bc\left[  \left(  x-1\right)  \partial_{x}+b\partial
_{b}\right]  ,\nonumber\\
E_{-\beta,-\gamma}  &  =\frac{1}{bc}\left[  x\left(  x-1\right)  \partial
_{x}+xa\partial_{a}-c\partial_{c}+1\right]  ,\nonumber\\
E_{\alpha\gamma}  &  =ac\left[  \left(  1-x\right)  \partial_{x}-a\partial
_{a}\right]  ,\nonumber\\
E_{-\alpha,-\gamma}  &  =\frac{1}{ac}\left[  x\left(  1-x\right)  \partial
_{x}-xb\partial_{b}+c\partial_{c}-1\right]  ,\nonumber\\
E_{\alpha\beta\gamma}  &  =abc\partial_{x},\nonumber\\
E_{-\alpha,-\beta,-\gamma}  &  =\frac{1}{abc}\left[  x\left(  x-1\right)
\partial_{x}-c\partial_{c}+xb\partial_{b}+xa\partial_{a}-x+1\right]
,\nonumber\\
J_{\alpha}  &  =a\partial_{a},\nonumber\\
J_{\beta}  &  =b\partial_{b},\nonumber\\
J_{\gamma}  &  =c\partial_{c}, \label{def}%
\end{align}
and calculate their operations on the basis functions \cite{sl4c,slkc}%

\begin{align}
E_{\alpha}f_{ac}^{b}\left(  \alpha;\beta;\gamma;x\right)   &  =\left(
\gamma-\alpha-1\right)  f_{ac}^{b}\left(  \alpha+1;\beta;\gamma;x\right)
,\nonumber\\
E_{\beta}f_{ac}^{b}\left(  \alpha;\beta;\gamma;x\right)   &  =\beta f_{ac}%
^{b}\left(  \alpha;\beta+1;\gamma;x\right)  ,\nonumber\\
E_{\gamma}f_{ac}^{b}\left(  \alpha;\beta;\gamma;x\right)   &  =\left(
\gamma-\beta\right)  f_{ac}^{b}\left(  \alpha;\beta;\gamma+1;x\right)
,\nonumber\\
E_{\beta\gamma}f_{ac}^{b}\left(  \alpha;\beta;\gamma;x\right)   &  =\beta
f_{ac}^{b}\left(  \alpha;\beta+1;\gamma+1;x\right)  ,\nonumber\\
E_{\alpha\gamma}f_{ac}^{b}\left(  \alpha;\beta;\gamma;x\right)   &  =\left(
\beta-\gamma\right)  f_{ac}^{b}\left(  \alpha+1;\beta;\gamma+1;x\right)
,\nonumber\\
E_{\alpha\beta\gamma}f_{ac}^{b}\left(  \alpha;\beta;\gamma;x\right)   &
=\beta f_{ac}^{b}\left(  \alpha+1;\beta+1;\gamma+1;x\right)  ,\nonumber\\
E_{-\alpha}f_{ac}^{b}\left(  \alpha;\beta;\gamma;x\right)   &  =\left(
\alpha-1\right)  f_{ac}^{b}\left(  \alpha-1;\beta;\gamma;x\right)
,\nonumber\\
E_{-\beta}f_{ac}^{b}\left(  \alpha;\beta;\gamma;x\right)   &  =\left(
\gamma-\beta\right)  f_{ac}^{b}\left(  \alpha;\beta-1;\gamma;x\right)
,\nonumber\\
E_{-\gamma}f_{ac}^{b}\left(  \alpha;\beta;\gamma;x\right)   &  =\left(
\alpha+1-\gamma\right)  f_{ac}^{b}\left(  \alpha;\beta;\gamma-1;x\right)
,\nonumber\\
E_{-\beta,-\gamma}f_{ac}^{b}\left(  \alpha;\beta;\gamma;x\right)   &  =\left(
\alpha-\gamma+1\right)  f_{ac}^{b}\left(  \alpha;\beta-1;\gamma-1;x\right)
,\nonumber\\
E_{-\alpha,-\gamma}f_{ac}^{b}\left(  \alpha;\beta;\gamma;x\right)   &
=\left(  \alpha-1\right)  f_{ac}^{b}\left(  \alpha-1;\beta;\gamma-1;x\right)
,\nonumber\\
E_{-\alpha,-\beta,-\gamma}f_{ac}^{b}\left(  \alpha;\beta;\gamma;x\right)   &
=\left(  -\alpha+1\right)  f_{ac}^{b}\left(  \alpha-1;\beta-1;\gamma
-1;x\right)  ,\nonumber\\
J_{\alpha}f_{ac}^{b}\left(  \alpha;\beta;\gamma;x\right)   &  =\alpha
f_{ac}^{b}\left(  \alpha;\beta;\gamma;x\right)  ,\nonumber\\
J_{\beta}f_{ac}^{b}\left(  \alpha;\beta;\gamma;x\right)   &  =\beta f_{ac}%
^{b}\left(  \alpha;\beta;\gamma;x\right)  ,\nonumber\\
J_{\gamma}f_{ac}^{b}\left(  \alpha;\beta;\gamma;x\right)   &  =\gamma
f_{ac}^{b}\left(  \alpha;\beta;\gamma;x\right)  . \label{op}%
\end{align}
Note, for example, that since $\beta$ is a nonpositive integer, the operation
by $E_{-\beta}$ will not be terminated as in the case of the finite
dimensional representation of a compact Lie group. Here the representation is
infinite dimensional. On the other hand, a simple calculation gives%

\begin{align*}
\left[  E_{\alpha},E_{-\alpha}\right]   &  =2J_{\alpha}-J_{\gamma},\\
\left[  E_{\beta},E_{-\beta}\right]   &  =2J_{\beta}-J_{\gamma},\\
\left[  E_{\gamma},E_{-\gamma}\right]   &  =2J_{\gamma}-\left(  J_{\alpha
}+J_{\beta}+1\right)  ,
\end{align*}
which suggest the Cartan subalgebra%
\begin{equation}
\left[  J_{\alpha},J_{\beta}\right]  =0,\left[  J_{\beta},J_{\gamma}\right]
=0,\left[  J_{\alpha},J_{\gamma}\right]  =0. \label{cartan}%
\end{equation}
Indeed, if we redefine%
\begin{align*}
J_{\alpha}^{\prime}  &  =J_{\alpha}-\frac{1}{2}J_{\gamma},\\
J_{\beta}^{\prime}  &  =J_{\beta}-\frac{1}{2}J_{\gamma},\\
J_{\gamma}^{\prime}  &  =J_{\gamma}-\frac{1}{2}\left(  J_{\alpha}+J_{\beta
}+1\right)  ,
\end{align*}
we find out that each of the triplets \cite{sl4c,slkc}%
\begin{align*}
&  \left\{  J^{+},J^{-},J^{0}\right\}  \equiv\left\{  E_{\alpha},E_{-\alpha
},J_{\alpha}^{\prime}\right\}  ,\left\{  E_{\beta},E_{-\beta},J_{\beta
}^{\prime}\right\}  ,\\
&  \left\{  E_{\gamma},E_{-\gamma},J_{\gamma}^{\prime}\right\}  ,\left\{
E_{\alpha,\beta,\gamma},E_{-\alpha,-\beta,-\gamma},J_{\alpha}^{\prime
}+J_{\beta}^{\prime}+J_{\gamma}^{\prime}\right\}  ,\\
&  \left\{  E_{\alpha\gamma},E_{-\alpha,-\gamma},J_{\alpha}^{\prime}%
+J_{\gamma}^{\prime}\right\}  ,\left\{  E_{\alpha\beta},E_{-\alpha,-\beta
},J_{\alpha}^{\prime}+J_{\beta}^{\prime}\right\}
\end{align*}
constitutes the well known commutation relations%
\begin{equation}
\left[  J^{0},J^{\pm}\right]  =\pm J^{\pm},\left[  J^{+},J^{-}\right]
=2J^{0}. \label{com}%
\end{equation}

In the following, we want to further relate the $SL(4,%
\mathbb{C}
)$ group to the recurrence relations of $F_{D}^{\left(  1\right)  }\left(
\alpha;\beta;\gamma;x\right)  $ or of the LSSA in Eq.(\ref{rel}). For our
purpose, there are $K+2=1+2=3$ recurrence relations among $F_{D}^{\left(
1\right)  }\left(  \alpha;\beta;\gamma;x\right)  $ or Gauss hypergeometry functions%

\begin{align}
\left(  \alpha-\beta\right)  F_{D}^{\left(  1\right)  }\left(  \alpha
;\beta;\gamma;x\right)  -\alpha F_{D}^{\left(  1\right)  }\left(
\alpha+1;\beta;\gamma;x\right)  +\beta F_{D}^{\left(  1\right)  }\left(
\alpha;\beta+1;\gamma;x\right)   &  =0,\label{R1}\\
\gamma F_{D}^{\left(  1\right)  }\left(  \alpha;\beta;\gamma;x\right)
-\left(  \gamma-\alpha\right)  F_{D}^{\left(  1\right)  }\left(  \alpha
;\beta;\gamma+1;x\right)  -\alpha F_{D}^{\left(  1\right)  }\left(
\alpha+1;\beta;\gamma+1;x\right)   &  =0,\label{R2}\\
\gamma F_{D}^{\left(  1\right)  }\left(  \alpha;\beta;\gamma;x\right)
+\gamma\left(  x-1\right)  F_{D}^{\left(  1\right)  }\left(  \alpha
;\beta+1;\gamma;x\right)  -\left(  \gamma-\alpha\right)  xF_{D}^{\left(
1\right)  }\left(  \alpha;\beta+1;\gamma+1;x\right)   &  =0. \label{R3}%
\end{align}
The three recurrence relations can be used to derive recurrence relations
among LSSA in Eq.(\ref{rel}).

In the following we will show that the three recurrence relations can be used
to reproduce the Cartan subalgebra and simple root system of the $SL(4,%
\mathbb{C}
)$ group with rank $3$. With the identification in Eq.(\ref{id}), the first
recurrence relation in Eq.(\ref{R1}) can be rewritten as
\begin{equation}
\frac{\left(  \alpha-\beta\right)  f_{ac}^{b}\left(  \alpha;\beta
;\gamma;x\right)  }{B\left(  \gamma-\alpha,\alpha\right)  a^{\alpha}b^{\beta
}c^{\gamma}}-\frac{\alpha f_{ac}^{b}\left(  \alpha+1;\beta;\gamma;x\right)
}{B\left(  \gamma-\alpha-1,\alpha+1\right)  a^{\alpha+1}b^{\beta}c^{\gamma}%
}+\frac{\beta f_{ac}^{b}\left(  \alpha;\beta+1;\gamma;x\right)  }{B\left(
\gamma-\alpha,\alpha\right)  a^{\alpha}b^{\beta+1}c^{\gamma}}=0.
\end{equation}
By using the identity%
\begin{equation}
B\left(  \gamma-\alpha-1,\alpha+1\right)  =\frac{\Gamma\left(  \gamma
-\alpha-1\right)  \Gamma\left(  \alpha+1\right)  }{\Gamma\left(
\gamma\right)  }=\frac{\alpha}{\gamma-\alpha-1}\frac{\Gamma\left(
\gamma-\alpha\right)  \Gamma\left(  \alpha\right)  }{\Gamma\left(
\gamma\right)  },
\end{equation}
the recurrence relation then becomes%
\begin{equation}
\left(  \alpha-\beta\right)  f_{ac}^{b}\left(  \alpha;\beta;\gamma;x\right)
-\frac{\alpha f_{ac}^{b}\left(  \alpha+1;\beta;\gamma;x\right)  }{\frac
{\alpha}{\gamma-\alpha-1}a}+\frac{\beta f_{ac}^{b}\left(  \alpha
;\beta+1;\gamma;x\right)  }{b}=0,
\end{equation}
or%
\begin{equation}
\left(  \alpha-\beta-\frac{E_{\alpha}}{a}+\frac{E_{\beta}}{b}\right)
f_{ac}^{b}\left(  \alpha;\beta;\gamma;x\right)  =0,
\end{equation}
which means%
\begin{equation}
\left[  \alpha-\beta-\left(  x\partial_{x}+a\partial_{a}\right)  +\left(
x\partial_{x}+b\partial_{b}\right)  \right]  f_{ac}^{b}\left(  \alpha
;\beta;\gamma;x\right)  =0,
\end{equation}
or%
\begin{equation}
\left[  \left(  \alpha-J_{\alpha}\right)  -\left(  \beta-J_{\beta}\right)
\right]  f_{ac}^{b}\left(  \alpha;\beta;\gamma;x\right)  =0. \label{R11}%
\end{equation}

Similarly for the second recurrence relation in Eq.(\ref{R2}), we obtain%
\begin{equation}
f_{ac}^{b}\left(  \alpha;\beta;\gamma;x\right)  -\frac{E_{\gamma}f_{ac}%
^{b}\left(  \alpha;\beta;\gamma;x\right)  }{c\left(  \gamma-\beta\right)
}-\frac{E_{\alpha\gamma}f_{ac}^{b}\left(  \alpha;\beta;\gamma;x\right)
}{ac\left(  \beta-\gamma\right)  }=0.
\end{equation}
After some calculations, we end up with%
\begin{equation}
\left[  \left(  \gamma-c\partial_{c}\right)  -\left(  \beta-b\partial
_{b}\right)  \right]  f_{ac}^{b}\left(  \alpha;\beta;\gamma;x\right)  =0,
\end{equation}
or%
\begin{equation}
\left[  \left(  \gamma-J_{\gamma}\right)  -\left(  \beta-J_{\beta}\right)
\right]  f_{ac}^{b}\left(  \alpha;\beta;\gamma;x\right)  =0. \label{R22}%
\end{equation}

Finally the third recurrence relation in Eq.(\ref{R3}) can be rewritten as
\begin{equation}
\gamma f_{ac}^{b}\left(  \alpha;\beta;\gamma;x\right)  +\frac{\gamma\left(
x-1\right)  E_{\beta}f_{ac}^{b}\left(  \alpha;\beta;\gamma;x\right)  }{b\beta
}-\frac{\left(  \gamma-\alpha\right)  xE_{\beta\gamma}f_{ac}^{b}\left(
\alpha;\beta;\gamma;x\right)  }{\frac{\gamma-\alpha}{\gamma}b\beta c}=0,
\end{equation}
which gives after some computation%
\begin{equation}
\left(  \beta-J_{\beta}\right)  f_{ac}^{b}\left(  \alpha;\beta;\gamma
;x\right)  =0. \label{R33}%
\end{equation}

It is easy to see that Eq.(\ref{R11}), Eq.(\ref{R22}) and Eq.(\ref{R33}) imply
the last three equations of Eq.(\ref{op}) or the Cartan subalgebra in
Eq.(\ref{cartan}) as expected.

In addition to the Cartan subalgebra, we need to derive the operations of the
$\{E_{\alpha},E_{\beta},E_{\gamma}\}$ from the recurrence relations. With the
operations of Cartan subalgebra and $\{E_{\alpha},E_{\beta},E_{\gamma}\}$, one
can reproduce the whole $SL(4,%
\mathbb{C}
)$ algebra. Note that the first recurrence relation in Eq.(\ref{R1}) can be
rewritten as%
\begin{equation}
\left(  \alpha-\beta\right)  f_{ac}^{b}\left(  \alpha;\beta;\gamma;x\right)
-\frac{\alpha f_{ac}^{b}\left(  \alpha+1;\beta;\gamma;x\right)  }{\frac
{\alpha}{\gamma-\alpha-1}a}+\frac{\beta f_{ac}^{b}\left(  \alpha
;\beta+1;\gamma;x\right)  }{b}=0,
\end{equation}
which means%
\begin{equation}
\left(  \alpha-\beta\right)  f_{ac}^{b}\left(  \alpha;\beta;\gamma;x\right)
-\frac{E_{a}f_{ac}^{b}\left(  \alpha;\beta;\gamma;x\right)  }{a}+\frac{\beta
f_{ac}^{b}\left(  \alpha;\beta+1;\gamma;x\right)  }{b}=0
\end{equation}
where we have used the operation of $E_{\alpha}$ in Eq.(\ref{op}). The next
step is to use the definition of $E_{\alpha}$ in Eq.(\ref{def}) to obtain%
\begin{equation}
\left(  \alpha-\beta-\frac{a\left(  x\partial_{x}+a\partial_{a}\right)  }%
{a}\right)  f_{ac}^{b}\left(  \alpha;\beta;\gamma;x\right)  =-\frac{\beta
f_{ac}^{b}\left(  \alpha;\beta+1;\gamma;x\right)  }{b},
\end{equation}
which implies%
\begin{equation}
\left[  b\left(  b\partial_{b}+x\partial_{x}\right)  \right]  f_{ac}%
^{b}\left(  \alpha;\beta;\gamma;x\right)  =E_{\beta}f_{ac}^{b}\left(
\alpha;\beta;\gamma;x\right)  =\beta f_{ac}^{b}\left(  \alpha;\beta
+1;\gamma;x\right)  \label{ok1}%
\end{equation}
where we have used the definition of $E_{\beta}$ in Eq.(\ref{def}).
Eq.(\ref{ok1}) is consistent with the operation of $E_{\beta}$ in Eq.(\ref{op}).

Similarly, we can check the operation of $E_{\alpha}$. Note that the first
recurrence relation in Eq.(\ref{R1}) can be rewritten as%
\begin{equation}
\left(  \alpha-\beta\right)  f_{ac}^{b}\left(  \alpha;\beta;\gamma;x\right)
-\frac{\left(  \gamma-\alpha-1\right)  f_{ac}^{b}\left(  \alpha+1;\beta
;\gamma;x\right)  }{a}+\frac{E_{\beta}f_{ac}^{b}\left(  \alpha;\beta
;\gamma;x\right)  }{b}=0
\end{equation}
where we have used the operation of $E_{\beta}$ in Eq.(\ref{op}). The next
step is to use the definition of $E_{\beta}$ in Eq.(\ref{def}) to obtain%
\begin{equation}
\left(  \alpha-\beta+\frac{b\left(  x\partial_{x}+b\partial_{b}\right)  }%
{b}\right)  f_{ac}^{b}\left(  \alpha;\beta;\gamma;x\right)  =\frac{\left(
\gamma-\alpha-1\right)  f_{ac}^{b}\left(  \alpha+1;\beta;\gamma;x\right)  }%
{a},
\end{equation}
which implies%
\begin{equation}
\left[  a\left(  a\partial_{a}+x\partial_{x}\right)  \right]  f_{ac}%
^{b}\left(  \alpha;\beta;\gamma;x\right)  =E_{\alpha}f_{ac}^{b}\left(
\alpha;\beta;\gamma;x\right)  =\left(  \gamma-\alpha-1\right)  f_{ac}%
^{b}\left(  \alpha+1;\beta;\gamma;x\right)  . \label{ok2}%
\end{equation}
where we have used the definition of $E_{\alpha}$ in Eq.(\ref{def}).
Eq.(\ref{ok2}) is consistent with the operation of $E_{\alpha}$ in
Eq.(\ref{op}).

Finally we check the operation of $E_{\gamma}$. Note that Eq.(\ref{R2}) can be
written as%
\begin{equation}
\frac{\gamma f_{ac}^{b}\left(  \alpha;\beta;\gamma;x\right)  }{B\left(
\gamma-\alpha,\alpha\right)  a^{\alpha}b^{\beta}c^{\gamma}}-\frac{\left(
\gamma-\alpha\right)  f_{ac}^{b}\left(  \alpha;\beta;\gamma+1;x\right)
}{\frac{\left(  \gamma-\alpha\right)  }{\gamma}B\left(  \gamma-\alpha
,\alpha\right)  a^{\alpha}b^{\beta}c^{\gamma+1}}-\frac{\alpha f_{ac}%
^{b}\left(  \alpha+1;\beta;\gamma+1;x\right)  }{\frac{\alpha}{\gamma}B\left(
\gamma-\alpha,\alpha\right)  a^{\alpha+1}b^{\beta}c^{\gamma+1}}=0,
\end{equation}
which gives%
\begin{equation}
f_{ac}^{b}\left(  \alpha;\beta;\gamma;x\right)  -\frac{f_{ac}^{b}\left(
\alpha;\beta;\gamma+1;x\right)  }{c}-\frac{f_{ac}^{b}\left(  \alpha
+1;\beta;\gamma+1;x\right)  }{ac}=0.
\end{equation}
The next step is to use the definition and operation of $E_{\alpha\gamma}$ to
obtain%
\begin{equation}
f_{ac}^{b}\left(  \alpha;\beta;\gamma;x\right)  -\frac{f_{ac}^{b}\left(
\alpha;\beta;\gamma+1;x\right)  }{c}-\frac{E_{\alpha\gamma}f_{ac}^{b}\left(
\alpha;\beta;\gamma;x\right)  }{ac\left(  \beta-\gamma\right)  }=0,\nonumber
\end{equation}
which gives%
\begin{equation}
f_{ac}^{b}\left(  \alpha;\beta;\gamma;x\right)  -\frac{ac\left[  \left(
1-x\right)  \partial_{x}-a\partial_{a}\right]  f_{ac}^{b}\left(  \alpha
;\beta;\gamma;x\right)  }{ac\left(  \beta-\gamma\right)  }=\frac{f_{ac}%
^{b}\left(  \alpha;\beta;\gamma+1;x\right)  }{c}.
\end{equation}
After some simple computation, we get%
\begin{align}
&  -c\left[  b\partial_{b}-c\partial_{c}-\left(  1-x\right)  \partial
_{x}+a\partial_{a}\right]  f_{ac}^{b}\left(  \alpha;\beta;\gamma;x\right)
\nonumber\\
&  =E_{\gamma}f_{ac}^{b}\left(  \alpha;\beta;\gamma;x\right)  =\left(
\gamma-\beta\right)  f_{ac}^{b}\left(  \alpha;\beta;\gamma+1;x\right)  .
\label{ok3}%
\end{align}
Eq.(\ref{ok3}) is consistent with the operation of $E_{\gamma}$ in
Eq.(\ref{op}).

We thus have shown that the \textit{extended }LSSA $f_{ac}^{b}\left(
\alpha;\beta;\gamma;x\right)  $ in Eq.(\ref{id}) with arbitrary $a$ and $c$
form an \textit{infinite} dimensional representation of the $SL(4,C)$ group.
Moreover, the $3$ recurrence relations among the LSSA can be used to reproduce
the Cartan subalgebra and simple root system of the $SL(4,%
\mathbb{C}
)$ group with rank $3$. The recurrence relations are thus equivalent to the
representation of the $SL(4,C)$ symmetry group.%

\setcounter{equation}{0}
\renewcommand{\theequation}{\arabic{section}.\arabic{equation}}%

\section{The General SL($K+3$,C) Symmetry}

To calculate the group representation of the LSSA for general $K$, we first
define \cite{slkc}
\begin{align}
&  f_{ac}^{b_{1}\cdots b_{K}}\left(  \alpha;\beta_{1},\cdots,\beta_{K}%
;\gamma;x_{1},\cdots,x_{K}\right) \nonumber\\
&  =B\left(  \gamma-\alpha,\alpha\right)  F_{D}^{\left(  K\right)  }\left(
\alpha;\beta_{1},\cdots,\beta_{K};\gamma;x_{1},\cdots,x_{K}\right)  a^{\alpha
}b_{1}^{\beta_{1}}\cdots b_{K}^{\beta_{K}}c^{\gamma}. \label{id2}%
\end{align}
Note that the LSSA in Eq.(\ref{st}) corresponds to the case $a=1=c$, and can
be written as%
\begin{equation}
A_{st}^{(r_{n}^{T},r_{m}^{P},r_{l}^{L})}=f_{11}^{-(n-1)!k_{3}^{T}%
,-(m-1)!k_{3}^{P},-(l-1)!k_{3}^{L}}\left(  -\frac{t}{2}-1;R_{n}^{T},R_{m}%
^{P},R_{l}^{L};\frac{u}{2}+2-N;\tilde{Z}_{n}^{T},\tilde{Z}_{m}^{P},\tilde
{Z}_{l}^{L}\right)  .
\end{equation}
It is possible to generalize the $SL(4,%
\mathbb{C}
)$ symmetry group for the $K=1$ case discussed in the previous section to the
general $SL(K+3,%
\mathbb{C}
)$ group. We first introduce the $(K+3)^{2}-1$ generators of $SL(K+3,C)$ group
($k=1,2,...K$) \cite{sl4c,slkc}%
\begin{align}
E^{\alpha}  &  =a\left(  \underset{j}{%
{\displaystyle\sum}
}x_{j}\partial_{j}+a\partial_{a}\right)  ,\nonumber\\
E^{\beta_{k}}  &  =b_{k}\left(  x_{k}\partial_{k}+b_{k}\partial_{b_{k}%
}\right)  ,\nonumber\\
E^{\gamma}  &  =c\left(  \underset{j}{%
{\displaystyle\sum}
}\left(  1-x_{j}\right)  \partial_{x_{j}}+c\partial_{c}-a\partial
_{a}-\underset{j}{%
{\displaystyle\sum}
}b_{j}\partial_{b_{j}}\right)  ,\nonumber\\
E^{\alpha\gamma}  &  =ac\left(  \underset{j}{%
{\displaystyle\sum}
}\left(  1-x_{j}\right)  \partial_{x_{j}}-a\partial_{a}\right)  ,\nonumber\\
E^{\beta_{k}\gamma}  &  =b_{k}c\left[  \left(  x_{k}-1\right)  \partial
_{x_{k}}+b_{k}\partial_{b_{k}}\right]  ,\nonumber\\
E^{\alpha\beta_{k}\gamma}  &  =ab_{k}c\partial_{x_{k}},\nonumber\\
E_{\alpha}  &  =\frac{1}{a}\left[  \underset{j}{%
{\displaystyle\sum}
}x_{j}\left(  1-x_{j}\right)  \partial_{x_{j}}+c\partial_{c}-a\partial
_{a}-\underset{j}{%
{\displaystyle\sum}
}x_{j}b_{j}\partial_{b_{j}}\right]  ,\nonumber\\
E_{\beta_{k}}  &  =\frac{1}{b_{k}}\left[  x_{k}\left(  1-x_{k}\right)
\partial_{x_{k}}+x_{k}\underset{j\neq k}{%
{\displaystyle\sum}
}\left(  1-x_{j}\right)  x_{j}\partial_{x_{j}}+c\partial_{c}-x_{k}%
a\partial_{a}-\underset{j}{%
{\displaystyle\sum}
}b_{j}\partial_{u_{j}}\right]  ,\nonumber\\
E_{\gamma}  &  =-\frac{1}{c}\left(  \underset{j}{%
{\displaystyle\sum}
}x_{j}\partial_{x_{j}}+c\partial_{c}-1\right)  ,\nonumber\\
E_{\alpha\gamma}  &  =\frac{1}{ac}\left[  \underset{j}{%
{\displaystyle\sum}
}x_{j}\left(  1-x_{j}\right)  \partial_{x_{j}}-\underset{j}{%
{\displaystyle\sum}
}x_{j}b_{j}\partial_{b_{j}}+c\partial_{c}-1\right]  ,\nonumber\\
E_{\beta_{k}\gamma}  &  =\frac{1}{b_{k}c}\left[  x_{k}\left(  x_{k}-1\right)
\partial_{x_{k}}+\underset{j\neq k}{%
{\displaystyle\sum}
}\left(  x_{j}-1\right)  x_{j}\partial_{x_{j}}+x_{k}a\partial_{a}%
-c\partial_{c}+1\right]  ,\nonumber\\
E_{\alpha\beta_{k}\gamma}  &  =\frac{1}{ab_{k}c}\left[  \underset{j}{%
{\displaystyle\sum}
}x_{j}\left(  x_{j}-1\right)  \partial_{x_{j}}-c\partial_{c}+x_{k}%
a\partial_{a}+\underset{j}{%
{\displaystyle\sum}
}x_{j}b_{j}\partial_{b_{j}}-x_{k}+1\right]  ,\nonumber\\
E_{\beta_{p}}^{\beta_{k}}  &  =\frac{b_{k}}{b_{p}}\left[  \left(  x_{k}%
-x_{p}\right)  \partial_{z_{k}}+b_{k}\partial_{b_{k}}\right]  ,(k\neq
p),\nonumber\\
J_{\alpha}  &  =a\partial_{a},\nonumber\\
J_{\beta_{k}}  &  =b_{k}\partial_{b_{k}},\nonumber\\
J_{\gamma}  &  =c\partial_{c}. \label{def2}%
\end{align}

Note that we have used the upper indices to denote the "raising operators" and
the lower indices to denote the "lowering operators". The number of generators
can be counted by the following way. There are $1$ $E^{\alpha}$, $K$
$E^{\beta_{k}}$, $1$ $E^{\gamma}$,$1$ $E^{\alpha\gamma}$,$K$ $E^{\beta
_{k}\gamma}$ and $K$ $E^{\alpha\beta_{k}\gamma}$ which sum up to $3K+3$
raising generators. There are also $3K+3$ lowering operators. In addition,
there are $K\left(  K-1\right)  $ $E_{\beta_{p}}^{\beta_{k}}$ and $K+2$
$\ $\ $\ J$ , the Cartan subalgebra. In sum, the total number of generators
are $2(3K+3)+K(K-1)+K+2=(K+3)^{2}-1$. It is straightforward to calculate the
operation of these generators on the basis functions ($k=1,2,...K$)
\cite{slkc}%

\begin{align}
E^{\alpha}f_{ac}^{b_{1}\cdots b_{K}}\left(  \alpha\right)   &  =\left(
\gamma-\alpha-1\right)  f_{ac}^{b_{1}\cdots b_{K}}\left(  \alpha+1\right)
,\nonumber\\
E^{\beta_{k}}f_{ac}^{b_{1}\cdots b_{K}}\left(  \beta_{k}\right)   &
=\beta_{k}f_{ac}^{b_{1}\cdots b_{K}}\left(  \beta_{k}+1\right)  ,\nonumber\\
E^{\gamma}f_{ac}^{b_{1}\cdots b_{K}}\left(  \gamma\right)   &  =\left(
\gamma-\underset{j}{%
{\displaystyle\sum}
}\beta_{j}\right)  f_{ac}^{b_{1}\cdots b_{K}}\left(  \gamma+1\right)
,\nonumber\\
E^{\alpha\gamma}f_{ac}^{b_{1}\cdots b_{K}}\left(  \alpha;\gamma\right)   &
=\left(  \underset{j}{%
{\displaystyle\sum}
}\beta_{j}-\gamma\right)  f_{ac}^{b_{1}\cdots b_{K}}\left(  \alpha
+1;\gamma+1\right)  ,\nonumber\\
E^{\beta_{k}\gamma}f_{ac}^{b_{1}\cdots b_{K}}\left(  \beta_{k};\gamma\right)
&  =\beta_{k}f_{ac}^{b_{1}\cdots b_{K}}\left(  \beta_{k}+1;\gamma+1\right)
,\nonumber\\
E^{\alpha\beta_{k}\gamma}f_{ac}^{b_{1}\cdots b_{K}}\left(  \alpha;\beta
_{k};\gamma\right)   &  =\beta_{k}f_{ac}^{b_{1}\cdots b_{K}}\left(
\alpha+1;\beta_{k}+1;\gamma+1\right)  ,\nonumber\\
E_{\alpha}f_{ac}^{b_{1}\cdots b_{K}}\left(  \alpha\right)   &  =\left(
\alpha-1\right)  f_{ac}^{b_{1}\cdots b_{K}}\left(  \alpha-1\right)
,\nonumber\\
E_{\beta_{k}}f_{ac}^{b_{1}\cdots b_{K}}\left(  \beta_{k}\right)   &  =\left(
\gamma-\underset{j}{%
{\displaystyle\sum}
}\beta_{j}\right)  f_{ac}^{b_{1}\cdots b_{K}}\left(  \beta_{k}-1\right)
,\nonumber\\
E_{\gamma}f_{ac}^{b_{1}\cdots b_{K}}\left(  \gamma\right)   &  =\left(
\alpha-\gamma+1\right)  f_{ac}^{b_{1}\cdots b_{K}}\left(  \gamma-1\right)
,\nonumber\\
E_{\alpha\gamma}f_{ac}^{b_{1}\cdots b_{K}}\left(  \alpha;\gamma\right)   &
=\left(  \alpha-1\right)  f_{ac}^{b_{1}\cdots b_{K}}\left(  \alpha
-1;\gamma-1\right)  ,\nonumber\\
E_{\beta_{k}\gamma}f_{ac}^{b_{1}\cdots b_{K}}\left(  \beta_{k};\gamma\right)
&  =\left(  \alpha-\gamma+1\right)  f_{ac}^{b_{1}\cdots b_{K}}\left(
\beta_{k}-1;\gamma-1\right)  ,\nonumber\\
E_{\alpha\beta_{k}\gamma}f_{ac}^{b_{1}\cdots b_{K}}\left(  \alpha;\beta
_{k};\gamma\right)   &  =\left(  1-\alpha\right)  f_{ac}^{b_{1}\cdots b_{K}%
}\left(  \alpha-1;\beta_{k}-1;\gamma-1\right)  ,\nonumber\\
E_{\beta_{p}}^{\beta_{k}}f_{ac}^{b_{1}\cdots b_{K}}\left(  \beta_{k};\beta
_{p}\right)   &  =\beta_{k}f_{ac}^{b_{1}\cdots b_{K}}\left(  \beta_{k}%
+1;\beta_{p}-1\right)  ,\nonumber\\
J_{\alpha}f_{ac}^{b_{1}\cdots b_{K}}\left(  \alpha;\beta_{k};\gamma\right)
&  =\alpha f_{ac}^{b_{1}\cdots b_{K}}\left(  \alpha;\beta_{k};\gamma\right)
,\nonumber\\
J_{\beta_{k}}f_{ac}^{b_{1}\cdots b_{K}}\left(  \alpha;\beta_{k};\gamma\right)
&  =\beta_{k}f_{ac}^{b_{1}\cdots b_{K}}\left(  \alpha;\beta_{k};\gamma\right)
,\nonumber\\
J_{\gamma}f_{ac}^{b_{1}\cdots b_{K}}\left(  \alpha;\beta_{k};\gamma\right)
&  =\gamma f_{ac}^{b_{1}\cdots b_{K}}\left(  \alpha;\beta_{k};\gamma\right)
\label{op2}%
\end{align}
where, for simplicity, we have omitted those arguments in $f_{ac}^{b_{1}\cdots
b_{K}}$ which remain the same after the operation. The commutation relations
of the $SL(K+3)$ Lie algebra can be calculated in the following way. In
addition to the Cartan subalgebra for the $K+2$ generators $\left\{
J_{\alpha},J_{\beta_{k}},J_{\gamma}\right\}  $, let's redefine%
\begin{align}
J_{\alpha}^{\prime}  &  =J_{\alpha}-\frac{1}{2}J_{\gamma},\nonumber\\
J_{\beta_{k}}^{\prime}  &  =J_{\beta_{k}}-\frac{1}{2}J_{\gamma}%
+\underset{j\neq k}{\sum}J_{\beta_{j}},\nonumber\\
J_{\gamma}^{\prime}  &  =J_{\gamma}-\frac{1}{2}\left(  J_{\alpha
}+\underset{j}{\sum}J_{\beta_{j}}+1\right)  .
\end{align}
One can show that each of the following triplets \cite{slkc}%
\begin{align}
&  \left\{  J^{+},J^{-},J^{0}\right\}  \equiv\left\{  E^{\alpha},E_{\alpha
},J_{\alpha}^{\prime}\right\}  ,\left\{  E^{\beta_{k}},E_{\beta_{k}}%
,J_{\beta_{k}}^{\prime}\right\}  ,\nonumber\\
&  \left\{  E^{\gamma},E_{\gamma},J_{\gamma}^{\prime}\right\}  ,\left\{
E^{\alpha\beta_{k}\gamma},E_{\alpha\beta_{k}\gamma},J_{\alpha}^{\prime
}+J_{\beta_{k}}^{\prime}+J_{\gamma}^{\prime}\right\}  ,\nonumber\\
&  \left\{  E^{\alpha\gamma},E_{\alpha\gamma},J_{\alpha}^{\prime}+J_{\gamma
}^{\prime}\right\}  ,\left\{  E^{\alpha\beta_{k}},E_{\alpha\beta_{k}%
},J_{\alpha}^{\prime}+J_{\beta_{k}}^{\prime}\right\}  ,\nonumber\\
&  \left\{  E_{\beta_{p}}^{\beta_{l}},E_{\beta_{l}}^{\beta_{p}},J_{\beta_{l}%
}^{\prime}-J_{\beta_{p}}^{\prime}\right\}  \label{33}%
\end{align}
satisfies the commutation relations in Eq.(\ref{com}).

There are $K+2$ fundamental recurrence relations among $F_{D}^{\left(
K\right)  }\left(  \alpha;\beta;\gamma;x\right)  $ or the Lauricella
functions. The three different kinds of recurrence relations are \cite{LLLY}%
\begin{equation}
\left(  \alpha-\underset{j}{\sum}\beta_{j}\right)  F_{D}^{\left(  K\right)
}-\alpha F_{D}^{(K)}\left(  \alpha+1\right)  +\underset{j}{%
{\displaystyle\sum}
}\beta_{j}F_{D}^{(K)}\left(  \beta_{j}+1\right)  =0, \label{RK1}%
\end{equation}

\begin{equation}
\gamma F_{D}^{(K)}-\left(  \gamma-\alpha\right)  F_{D}^{(K)}\left(
\gamma+1\right)  -\alpha F_{D}^{(K)}\left(  \alpha+1;\gamma+1\right)  =0,
\label{RK2}%
\end{equation}
and%
\begin{equation}
\gamma F_{D}^{(K)}+\gamma(x_{m}-1)F_{D}^{(K)}\left(  \beta_{m}+1\right)
+\left(  \alpha-\gamma\right)  x_{m}F_{D}^{(K)}\left(  \beta_{m}%
+1;\gamma+1\right)  =0 \label{RK3}%
\end{equation}
where $m=1,2,...K.$ The three types of recurrence relations can be used to
derive recurrence relations among LSSA and reduce the number of
independent\ LSSA from $\infty$ down to $1$ \cite{LLLY}.

In the following we will show that the three types of recurrence relations
above imply the Cartan subalgebra of the $SL(K+3,%
\mathbb{C}
)$ group with rank $K+2$. With the identification in Eq.(\ref{id2}), the first
type of recurrence relation in Eq.(\ref{RK1}) can be rewritten as%
\begin{equation}
\left(  \alpha-\underset{j}{\sum}\beta_{j}\right)  f_{ac}^{b_{1}\cdots b_{K}%
}-\frac{E^{\alpha}f_{ac}^{b_{1}\cdots b_{K}}\left(  \alpha\right)  }%
{a}+\underset{j}{%
{\displaystyle\sum}
}\frac{E^{\beta_{j}}f_{ac}^{b_{1}\cdots b_{K}}\left(  \beta_{j}\right)
}{b_{j}}=0,
\end{equation}
which gives%
\begin{equation}
\left(  \alpha-\underset{j}{\sum}\beta_{j}\right)  f_{ac}^{b_{1}\cdots b_{K}%
}-\left(  \underset{j}{%
{\displaystyle\sum}
}x_{j}\partial_{j}+a\partial_{a}\right)  f_{ac}^{b_{1}\cdots b_{K}%
}+\underset{j}{%
{\displaystyle\sum}
}\left(  x_{j}\partial_{j}+b_{j}\partial_{b_{j}}\right)  f_{ac}^{b_{1}\cdots
b_{K}}=0
\end{equation}
or%
\begin{equation}
\left[  \left(  \alpha-a\partial_{a}\right)  +\underset{j}{\sum}\left(
\beta_{j}-b_{j}\partial_{b_{j}}\right)  \right]  f_{ac}^{b_{1}\cdots b_{K}}=0,
\label{car1}%
\end{equation}
which means%
\begin{equation}
\left[  \left(  \alpha-J_{\alpha}\right)  +\underset{j}{\sum}\left(  \beta
_{j}-J_{\beta_{j}}\right)  \right]  f_{ac}^{b_{1}\cdots b_{K}}=0.
\label{car11}%
\end{equation}

The second type of recurrence relation in Eq.(\ref{RK2}) can be rewritten as%
\begin{equation}
f_{ac}^{b_{1}\cdots b_{K}}-\frac{E^{\gamma}f_{ac}^{b_{1}\cdots b_{K}}\left(
\gamma\right)  }{c\left(  \gamma-\underset{j}{%
{\displaystyle\sum}
}\beta_{j}\right)  }-\frac{E^{\alpha\gamma}f_{ac}^{b_{1}\cdots b_{K}}\left(
\alpha;\gamma\right)  }{ac\left(  \underset{j}{%
{\displaystyle\sum}
}\beta_{j}-\gamma\right)  }=0,
\end{equation}
which gives%
\begin{equation}
\left[
\begin{array}
[c]{c}%
\gamma-\underset{j}{%
{\displaystyle\sum}
}\beta_{j}-\left(  \underset{j}{%
{\displaystyle\sum}
}\left(  1-x_{j}\right)  \partial_{x_{j}}+c\partial_{c}-a\partial
_{a}-\underset{j}{%
{\displaystyle\sum}
}b_{j}\partial_{b_{j}}\right) \\
+\left(  \underset{j}{%
{\displaystyle\sum}
}\left(  1-x_{j}\right)  \partial_{x_{j}}-a\partial_{a}\right)
\end{array}
\right]  f_{ac}^{b_{1}\cdots b_{K}}=0
\end{equation}
or%
\begin{equation}
\left[  \left(  \gamma-c\partial_{c}\right)  -\underset{j}{%
{\displaystyle\sum}
}\left(  \beta_{j}-b_{j}\partial_{b_{j}}\right)  \right]  f_{ac}^{b_{1}\cdots
b_{K}}=0. \label{car2}%
\end{equation}
Eq.(\ref{car2}) can be written as%
\begin{equation}
\left[  \left(  \gamma-J_{\gamma}\right)  -\underset{j}{%
{\displaystyle\sum}
}\left(  \beta_{j}-J_{\beta_{j}}\right)  \right]  f_{ac}^{b_{1}\cdots b_{K}%
}=0. \label{car22}%
\end{equation}

The third type of recurrence relation in Eq.(\ref{RK3}) can be rewritten as
($m=1,2,...K$)%
\begin{equation}
f_{ac}^{b_{1}\cdots b_{K}}+\frac{(x_{m}-1)E^{\beta_{m}}f_{ac}^{b_{1}\cdots
b_{K}}}{b_{m}\beta_{m}}-\frac{x_{m}E^{\beta_{m}\gamma}f_{ac}^{b_{1}\cdots
b_{K}}}{b_{m}c\beta_{m}}=0,
\end{equation}
which gives%
\begin{equation}
\beta_{m}f_{ac}^{b_{1}\cdots b_{K}}+(x_{m}-1)\left(  x_{m}\partial_{m}%
+b_{m}\partial_{b_{m}}\right)  f_{ac}^{b_{1}\cdots b_{K}}-x_{m}\left[  \left(
x_{m}-1\right)  \partial_{x_{m}}+b_{m}\partial_{b_{m}}\right]  f_{ac}%
^{b_{1}\cdots b_{K}}=0
\end{equation}
or%
\begin{equation}
\left(  \beta_{m}-b_{m}\partial_{b_{m}}\right)  f_{ac}^{b_{1}\cdots b_{K}}=0.
\label{car3}%
\end{equation}
In the above calculation, we have used the definition and operation of
$E^{\beta_{m}\gamma}$ in Eq.(\ref{def2}) and Eq.(\ref{op2}), respectively.

Eq.(\ref{car3}) can be written as%
\begin{equation}
\left(  \beta_{m}-J_{\beta_{m}}\right)  f_{ac}^{b_{1}\cdots b_{K}%
}=0,m=1,2,...K. \label{car33}%
\end{equation}

It is important to see that Eq.(\ref{car11}), Eq.(\ref{car22}) and
Eq.(\ref{car33}) imply the last three equations of Eq.(\ref{op2}) or the
Cartan subalgebra of $SL(K+3,C)$ as expected.

In addition to the Cartan subalgebra, we need to derive the operations of the
$\{E^{\alpha},E^{\beta_{k}},E^{\gamma}\}$ from the recurrence relations. With
the operations of Cartan subalgebra and $\{E^{\alpha},E^{\beta_{k}},E^{\gamma
}\}$, one can reproduce the whole $SL(K+3,%
\mathbb{C}
)$ algebra. The calculations of $E^{\alpha}$ and $E^{\gamma}$ are
straightforward and are similar to the case of $SL(4,%
\mathbb{C}
)$ in the previous section. Here we present only the calculation of
$E^{\beta_{k}}$. The recurrence relation in Eq.(\ref{RK1}) can be rewritten as%
\begin{equation}
\left(  \alpha-\underset{j}{\sum}\beta_{j}\right)  f_{ac}^{b_{1}\cdots b_{K}%
}-\frac{E^{\alpha}f_{ac}^{b_{1}\cdots b_{K}}\left(  \alpha\right)  }%
{a}+\underset{j\neq k}{%
{\displaystyle\sum}
}\frac{E^{\beta_{j}}f_{ac}^{b_{1}\cdots b_{K}}\left(  \beta_{j}\right)
}{b_{j}}+\frac{\beta_{k}f_{ac}^{b_{1}\cdots b_{K}}\left(  \beta_{k}+1\right)
}{b_{k}}=0.
\end{equation}
After operation of $E^{\beta_{j}}$, we obtain%
\begin{align}
&  \left(  \alpha-\underset{j}{\sum}\beta_{j}\right)  f_{ac}^{b_{1}\cdots
b_{K}}-\left(  \underset{j}{%
{\displaystyle\sum}
}x_{j}\partial_{j}+a\partial_{a}\right)  f_{ac}^{b_{1}\cdots b_{K}}\nonumber\\
+\underset{j\neq k}{%
{\displaystyle\sum}
}\left(  x_{j}\partial_{j}+b_{j}\partial_{b_{j}}\right)  f_{ac}^{b_{1}\cdots
b_{K}}  &  =\frac{-\beta_{k}f_{ac}^{b_{1}\cdots b_{K}}\left(  \beta
_{k}+1\right)  }{b_{k}},
\end{align}
which gives the consistent result%
\begin{equation}
b_{k}\left(  b_{k}\partial_{b_{k}}+x_{k}\partial_{k}\right)  f_{ac}%
^{b_{1}\cdots b_{K}}\left(  \beta_{k}\right)  =E^{\beta_{k}}f_{ac}%
^{b_{1}\cdots b_{K}}=\beta_{k}f_{ac}^{b_{1}\cdots b_{K}}\left(  \beta
_{k}+1\right)  ,k=1,2,...K.
\end{equation}
In the above calculation, we have used the definitions and operations of
$E^{\beta_{k}}$ and $E^{\alpha}$ in Eq.(\ref{def2}) and Eq.(\ref{op2}), respectively.

The $K+2$ equations in Eq.(\ref{car11}), Eq.(\ref{car22}) and Eq.(\ref{car33})
together with $K+2$ equations for the operations $\{E^{\alpha},E^{\beta_{k}%
},E^{\gamma}\}$ are equivalent to the Cartan subalgebra and the simple root
system of $SL(K+3,C)$ with rank $K+2.$ With the Cartan subalgebra and the
simple roots, one can easily write down the whole Lie algebra of the
$SL(K+3,C)$ group. So one can construct the Lie algebra from the recurrence
relations and vice versa.

In the previous publication, it was shown that \cite{LLLY} the $K+2$
recurrence relations among $F_{D}^{(K)}$ can be used to derive recurrence
relations among LSSA and reduce the number of independent\ LSSA from $\infty$
down to $1$. We conclude that the $SL(K+3,%
\mathbb{C}
)$ group can be used to derive infinite number of recurrence relations among
LSSA, and one can solve all the LSSA and express them in terms of one amplitude.

Finally, in addition to Eq.(\ref{33}), there is a simple way to write down the
Lie algebra commutation relations of $SL(K+3,C),$namely \cite{slkc}
\begin{equation}
\left[  \mathcal{E}_{ij},\mathcal{E}_{kl}\right]  =\delta_{jk}\mathcal{E}%
_{il}-\delta_{li}\mathcal{E}_{kj}%
\end{equation}
with the identifications%
\begin{align}
E^{\alpha}  &  =\mathcal{E}_{12},E_{\alpha}=\mathcal{E}_{21},E^{\beta_{k}%
}=\mathcal{E}_{k+3,3},E_{\beta}=\mathcal{E}_{3,k+3},\nonumber\\
E^{\gamma}  &  =\mathcal{E}_{31},E_{\gamma}=\mathcal{E}_{13},E^{\alpha\gamma
}=\mathcal{E}_{32},E_{\alpha\gamma}=\mathcal{E}_{23},\nonumber\\
E^{\beta_{k}\gamma}  &  =-\mathcal{E}_{k+3,1},E_{\beta_{k}\gamma}%
=-\mathcal{E}_{1,k+3},E_{\alpha\beta_{k}\gamma}=-\mathcal{E}_{k+3,2}%
,\nonumber\\
E_{\alpha\beta_{k}\gamma}  &  =-\mathcal{E}_{2,k+3},J_{\alpha}^{\prime}%
=\frac{1}{2}\left(  \mathcal{E}_{11}-\mathcal{E}_{22}\right)  ,J_{\beta_{k}%
}^{\prime}=\frac{1}{2}\left(  \mathcal{E}_{k+3,k+3}-\mathcal{E}_{33}\right)
,J_{\gamma}^{\prime}=\frac{1}{2}\left(  \mathcal{E}_{33}-\mathcal{E}%
_{11}\right)  .
\end{align}
%

\setcounter{equation}{0}
\renewcommand{\theequation}{\arabic{section}.\arabic{equation}}%

\section{Conclusion and Discussion}

In this paper, we point out that the exact LSSA in the $26D$ open bosonic
string theory can be expressed in terms of the basis functions in the infinite
dimensional representation space $V$ of the $SL(K+3,%
\mathbb{C}
)$ group which contains the $SO(2,1)$ spacetime Lorentz group. In addition, we
find that the $K+2$ recurrence relations among the LSSA can be used to
reproduce the Cartan subalgebra and simple root system of the $SL(K+3,%
\mathbb{C}
)$ group with rank $K+2$. Thus the recurrence relations are equivalent to the
representation of $SL(K+3,%
\mathbb{C}
)$ group of the LSSA. As a result, the $SL(K+3,%
\mathbb{C}
)$ group can be used to solve all the LSSA and express them in terms of one
amplitude \cite{LLLY} . As an application in the hard scattering limit, the
$SL(K+3,%
\mathbb{C}
)$ group can be used to directly prove Gross conjecture
\cite{GM,Gross,GrossManes}, which was previously corrected and proved by the
method of decoupling of zero norm states \cite{ChanLee1,ChanLee2,CHL,PRL,
CHLTY,susy,ZNS1}.

There are some special properties in the $SL(K+3,%
\mathbb{C}
)$ group representation of the LSSA, which make it different from the usual
symmetry group representation of a physical system. First, the set of LSSA
does not fill up the whole representation space $V$. For example, for states
$f_{ac}^{b_{1}\cdots b_{K}}\left(  \alpha;\beta_{1},\cdots,\beta_{K}%
;\gamma;x_{1},\cdots,x_{K}\right)  $ in $V$ with $a\neq1$ or $c\neq1$, they
are not LSSA.

Indeed, there are more states in $V$ with $K\geq2$ which are not LSSA either.
We give one example in the following. For $K=2$ there are six type of LSSA
$(\omega=-1)$%
\begin{align}
(\alpha_{-1}^{T})^{p_{1}}(\alpha_{-1}^{P})^{q_{1}}\text{,}F_{D}^{(2)}%
(a,-p_{1},-q_{1},c-p_{1}-q_{1},1,\left[  \tilde{z}_{1}^{P}\right]  )\text{,}N
&  =p_{1}+q_{1},\\
(\alpha_{-1}^{T})^{p_{1}}(\alpha_{-1}^{L})^{r_{1}}\text{,}F_{D}^{(2)}%
(a,-p_{1},-r_{1},c-p_{1}-r_{1},1,\left[  \tilde{z}_{1}^{L}\right]  )\text{,}N
&  =p_{1}+r_{1},\\
(\alpha_{-1}^{P})^{q_{1}}(\alpha_{-1}^{L})^{r_{1}}\text{,}F_{D}^{(2)}%
(a,-q_{1},-r_{1},c-q_{1}-r_{1},\left[  \tilde{z}_{1}^{P}\right]  ,\left[
\tilde{z}_{1}^{L}\right]  )\text{,}N  &  =q_{1}+r_{1},\\
(\alpha_{-2}^{T})^{p_{2}}\text{ , }F_{D}^{(2)}(a,-p_{2},-p_{2},c-2p_{2}%
,1,1)\text{ , }N  &  =2p_{2},\label{2a}\\
(\alpha_{-2}^{P})^{q_{2}}\text{ , }F_{D}^{(2)}(a,-q_{2},-q_{2},c-2q_{2}%
,1-z_{2}^{P},1-\omega z_{2}^{P}),N  &  =2q_{2},\label{2b}\\
(\alpha_{-2}^{L})^{r_{2}}\text{ , }F_{D}^{(2)}(a,-r_{2},-r_{2},c-2r_{2}%
,1-z_{2}^{L},1-\omega z_{2}^{L}),N  &  =2r_{2}. \label{2c}%
\end{align}
One can show that the states obtained from operation by $E_{\beta}$ on either
states in Eq.(\ref{2a}) to Eq.(\ref{2c}) are not LSSA. However, all states in
$V$ including those "auxiliary states" which are not LSSA can be exactly
solved by recurrence relations or the $SL(K+3,%
\mathbb{C}
)$ group and express them in terms of one amplitude. These "auxiliary states"
and states with $a\neq1$ or $c\neq1$ in $V$ may represent other SSA, e.g. SSA
of two tachyon and two arbitrary string states etc. Work in this direction is
in progress.

\subsection{Discussion}

In quantum field theory (QFT), one usually considers interactions with up to
four point couplings. It is the symmetry principle which fixs the forms of the
couplings. In addition, symmetry can be used to derive Ward identities and
relate different couplings. In string theory, on the contrary, one is given a
set of rules through quantum consistency of extended string to calculate
perturbative on-shell SSA. Moreover, instead of up to four point couplings in
QFT, one encounters $n$-point couplings with arbitrary $n$ in string theory.

In this paper, among these $n$-point couplings, we have calculated a subset of
four point SSA, namely, three tachyons and one arbitrary string states, and
the $SL(K+3,%
\mathbb{C}
)$ group was shown to be associated with these SSA. Presumably, in general,
there exists a huge symmetry group$\ G$ which is associated with the general
four point SSA%
\begin{equation}
<V_{1}V_{2}V_{3}V_{4}> \label{vv}%
\end{equation}
where $V_{j}$ , $j=1,2,3,4$ can be arbitrary string states. In particular, for
the case of $V_{1}$, $V_{3}$ and $V_{4}$ to be tachyons, one recovers in
Eq.(\ref{vv}) the LSSA of three tachyons and one arbitrary string state with
$SL(K+3,%
\mathbb{C}
)$ symmetry. It is reasonable to believe that the $SL(K+3,%
\mathbb{C}
)$ group forms a subgroup of the bigger group $G$.

In sum, although we have considered only a special class of SSA in this paper,
we conjecture that the $SL(K+3,%
\mathbb{C}
)$ symmetry group we discovered for the LSSA should survive for general string
interactions, at least for the symmetry structure of general four point SSA.

\noindent

\begin{acknowledgments}
We would like to thank H. Kawai for his suggestion in the early stage of this
work. This work is supported in part by the Ministry of Science and Technology
(MoST) and S.T. Yau center of National Chiao Tung University (NCTU), Taiwan.
\end{acknowledgments}

\end{document}